\documentclass[10pt,
superscriptaddress,
nofootinbib,
bibnotes,
amsmath,amssymb,
aps,
pra,
]{revtex4-2}

\usepackage[utf8]{inputenc}
\usepackage{siunitx}
\usepackage[english]{babel}			
\usepackage{graphicx}				
\usepackage{amssymb,amsmath,mathtools, bbold}		
\usepackage{url}					
\usepackage{caption}

\usepackage{braket}

\usepackage{tikz}
\usetikzlibrary{positioning}
\usetikzlibrary{calc}

\usepackage{bm}
\usepackage{hyperref}
\hypersetup{colorlinks=true,allcolors=blue}
\usepackage{lineno}

\DeclareCaptionLabelFormat{custom}{#1 \textbf{#2}}
\AtBeginDocument{\RenewCommandCopy\qty\SI}

\preprint{APS/123-QED}

\begin{document}

\makeatletter
\patchcmd{\frontmatter@RRAP@format}{(}{}{}{}
\patchcmd{\frontmatter@RRAP@format}{)}{}{}{}
\renewcommand\Dated@name{}
\makeatother
\date{Date: \today \\ \phantom{XXX} E-mail: stefan.frick@uibk.ac.at}

\title{Deterministic nonlinear domain engineering algorithm with variable sub-coherence-length domains}

\author{Michael Schlosser} 

\affiliation{Institut für Experimentalphysik, Universität Innsbruck, Technikerstr. 25, 6020 Innsbruck, Austria}

\author{Tommaso Faleo} 
\affiliation{Institut für Experimentalphysik, Universität Innsbruck, Technikerstr. 25, 6020 Innsbruck, Austria}

\author{Gregor Weihs} 
\affiliation{Institut für Experimentalphysik, Universität Innsbruck, Technikerstr. 25, 6020 Innsbruck, Austria}

\author{Robert Keil} 
\affiliation{Institut für Experimentalphysik, Universität Innsbruck, Technikerstr. 25, 6020 Innsbruck, Austria}

\author{Stefan Frick} 
\email{stefan.frick@uibk.ac.at}
\affiliation{Institut für Experimentalphysik, Universität Innsbruck, Technikerstr. 25, 6020 Innsbruck, Austria}

\begin{abstract}

We present a new and improved deterministic algorithm that calculates the poling domains for the implementation of arbitrary phase-matching functions (PMF) in quasi-phase-matched nonlinear crystals for parametric down-conversion photon sources. In contrast to previous work, the algorithm presented here makes full use of the limits of the fabrication process, such as the minimum feature size and resolution of the lithographic process underpinning the poling procedure. This minimizes the tracking error to the target electric field and improves the accuracy in reaching the desired PMF. We give a tutorial on how to calculate the target function for arbitrary PMFs and run the algorithm. We also demonstrate the procedure on a relevant application in quantum sensing, where we use a quadratic phase on the PMF to implement a broadband photon spectrum while maintaining high nonlinearity across the crystal. The designed crystal is experimentally tested, and the result matches expectations.
\end{abstract}
\maketitle
\section*{Introduction}\label{sec:intro}
The development of versatile, bright and practical sources for quantum states of light is of great interest for photonic quantum technologies~\cite{O'Brien2009,Flamini2019}. 
Single-photons or entangled photon pairs, to name types of such quantum states, are used in technologies such as quantum communication~\cite{Bennett1993,Couteau2023_1}, cryptography~\cite{Gisin2002}, metrology~\cite{Kok2016,Lyons2018,Couteau2023_2}, or computing~\cite{Knill2001}. 

Popular sources for these states include, for example, quantum dots~\cite{Senellart2017}, colour centres in diamond~\cite{Roch2006}, or nonlinear optics such as spontaneous parametric down-conversion (SPDC)~\cite{Burnham1970} or spontaneous four-wave mixing~\cite{chen2005}. 
SPDC in particular is very versatile and has been successfully used to achieve high spectral purities~\cite{Mosley2008,Branczyk2011,Pickston21,faleo2025} and to generate entangled states in various degrees of freedom, such as polarization~\cite{Kwiat1995,Fedrizzi2007}, energy-time~\cite{Ekert1992,Brendel1999}, frequency-bins~\cite{Kaneda2019,Morrison2022}, pulse modes \cite{Ansari2018,Graffitti2020} or combinations thereof \cite{Reimer2019, Imany2019,Chiriano2023,Lu2023,faleo2026}. 
Widely used materials for this process are non-inversion symmetric crystals such as potassium titanyl phosphate (KTP) or lithium niobate. 
By alternating the orientation of the crystals during the growth process via electric poling, the nonlinear susceptibility of the material can be controlled and quasi-phase-matching can be achieved via constructive interference of the down-converted fields from subsequent domains \cite{Hum2007}. 
The most basic poling structures achieving quasi-phase-matching consist of a fixed domain width that equals the coherence length of the nonlinear process. 
These simple structures have a fixed period, a \SI{50}{\percent} duty cycle, and produce a (often undesired) $\mathrm{sinc}$-shaped phase-matching function (PMF) . The $\mathrm{sinc}$-shape arises from the Fourier transform of the constant effective nonlinearity within the nonlinear medium. 
Different PMFs can be realized by varying the duty cycle and period to alter the effective nonlinearity profile along the medium's propagation axis. 
Such domain engineering is a flexible procedure to create 
photonic quantum states with engineered spectral and temporal properties~\cite{Branczyk2011,Weiss2025}. 
State-of-the-art algorithms use domains of predetermined width and vary their poling orientation~\cite{Tambasco2016,Graffitti2017}.
Analytical approaches that obtain a poling pattern directly from the desired nonlinear profile have also been proposed and tested~\cite{Shiloh_12,Leshem_14,Chekhova2018}.
Applications of these algorithms range from quantum communication~\cite{Ho2022,Graffitti2020_jan}, over the generation of tailored pulse-mode entanglement useful for quantum information~\cite{Graffitti2020}, to the generation of spectrally pure states \cite{Graffitti_18_may,Pickston21,faleo2025} or states with tailored spectral correlations \cite{Shukhin2024}.

Here, we present an improved algorithm that sets the width and orientation of each domain individually, utilizing the full manufacturing potential of nonlinear crystals and, thereby, achieving smaller deviations between the target and generated down-converted fields compared to previous algorithms. 
We provide a practical guide on how to apply it for a large variety of custom requirements. 
As a specific application example, we demonstrate step-by-step the design of a degenerate SPDC-pair source with Gaussian spectra of broad bandwidth and high brightness, and experimentally validate it. 

Section~\ref{sec:dom_eng} introduces nonlinear domain engineering and presents a new algorithm, which deterministically finds poling structures with increased performance. In section~\ref{sec:bright_broad_source}, this algorithm is used to produce a novel design for a nonlinear crystal used as a bright and broadband photon pair source for quantum sensing. The performance of this crystal is experimentally verified in section~\ref{sec:experimental_results}. 
Furthermore, a detailed step-by-step tutorial is presented in section~\ref{sec:setpbystep}, guiding researchers through the domain engineering process. 
Section~\ref{sec:improvments} discusses further improvements to the algorithm, which help in tracking the PMF over large frequency ranges.

\section{Domain engineering}\label{sec:dom_eng}

 Domain engineering finds poling structures of domains inside nonlinear media. Typical poling structures consist of a fixed domain width that equals the coherence length of the nonlinear process, thus enabling quasi-phase-matching.

\subsection{Quasi-phase-matching}\label{sec:Quasi-phase-matching}

Quasi-phase-matching enables nonlinear optical processes for combinations of crystal materials and wavelengths, which are inaccessible to conventional methods such as birefringent phase-matching~\cite{Hum2007}. 
For three-wave-mixing processes (such as SPDC), the phase-mismatch is $\Delta k \, = \, k_\text{p} - k_\text{s} - k_\text{i}$, where $k_j$, with $j \in \{\text{p,s,i} \}$, is the wavenumber of the pump, signal and idler photons, respectively and $k_j=\frac{2\pi n_j}{\lambda_j}$, with $\lambda_j$($n_j$) denoting the wavelength (refractive index) of the respective wave. 
The coherence length $l_{\text{c}}$ of the nonlinear process is given by
\begin{equation}
  \label{eq:lc_def}
  l_{\text{c}}  =  \frac{\pi}{\Delta k}.
\end{equation}
It defines the length after which the converted waves (signal or idler) begin to interfere destructively with themselves. 
By flipping the lattice orientation within the nonlinear crystal every coherence length, an effective $\pi$-phase-shift is introduced. 
This allows for constructive interference of the nonlinear process and yields an effective nonlinearity profile $g(z)={\chi^{(2)}_\text{eff}(z)}/{\chi^{(2)}_{\text{bulk}}}$, where $\chi^{(2)}_{\text{eff}}$ is the effective nonlinear coefficient of second order and $\chi^{(2)}_{\text{bulk}}$ is the nonlinear coefficient of the bulk material~\cite{Frick2023}. 
The drawback of quasi-phase-matching is that it has a maximum achievable effective nonlinearity of $\frac{2}{\pi}<1$.
However, quasi-phase-matching allows for much greater flexibility than other phase-matching methods. 
By carefully designing the width and sign of each nonlinear domain, a process known as domain engineering, the tailoring of the biphoton spectral amplitude to specific applications becomes possible. 
This is, for example, the reason why heralded single-photon sources with domain engineered nonlinear crystals exhibit record purities~\cite{Pickston21,faleo2025}. The PMF
\begin{equation}\label{eq:PMF}
    \phi \left( \Delta k \right )=\int_{-\infty}^{+\infty}g(z) \, e^{i \, \Delta k \, z}dz
\end{equation}
is the Fourier transform of the effective nonlinearity profile $g(z)$, which alternates its sign between domains. 
Consequently, varying the effective nonlinearity along the crystal coordinate $z$, enables tuning of the nonlinearity profile and PMF. 
Finding an optimal poling for a desired PMF is not trivial. 
Since $\chi^{(2)}_{\text{eff}}$ is constant within each domain, the desired nonlinearity profile can only be approximated by a series of step functions with finite widths. 
Moreover, practical considerations such as the minimum feature size and resolution of the lithographic process used in the manufacturing of the poling structure play a crucial role. 
Deterministic algorithms find working poling structures with relative ease, and the theoretical background is well documented in~\cite{Tambasco2016,Graffitti2017}.

\subsection{Domain engineering algorithms}\label{sec:doma-engin-algor}

First algorithms used simulated annealing~\cite{Reid2003,Dosseva2016} and other analytical methods to find poling structures of nonlinear crystals~\cite{Branczyk2011,Shiloh_12,Leshem_14,Dixon2013}, with further improvements found by gradient descent approaches~\cite{Cui2019}.
Deterministic tracking algorithms have since been demonstrated, which dramatically reduce the computational effort compared to their predecessors~\cite{Tambasco2016,Graffitti2017}.
These greedy optimization algorithms set the poling structure from the beginning to the end of the nonlinear crystal by tracking the complex electric field amplitude of the generated (signal or idler) waves and try to match it to the desired field.  
In principle, the desired field amplitude profile 
  \begin{equation}\label{eq:target_field}
    A_{\text{target}}(z)=\int_{0}^{z} g_{\text{target}}(\tilde{z}) \,  d\tilde{z}.
    \end{equation}
is calculated from an integral over the desired nonlinearity profile $g_{\text{target}}(z)$, which is, in turn, given by the inverse Fourier transform of the desired PMF \cite{Tambasco2016}
  \begin{equation}\label{eq:target_nonlinear}
    g_{\text{target}}(z) \propto \int_{-\infty}^{+\infty} \phi_{\text{target}}({\Delta k}) \, e^{-i\,z\,\Delta k} \,d\Delta k.
    \end{equation}
Note that $g_{\text{target}}(z)$ is continuous and complex-valued, with no additional constraints. 
This approach is much more efficient than calculating the PMF in every step of an annealing algorithm. 
For example, a field amplitude following an error function is desired, when the PMF should be a Gaussian curve. 
The resulting nonlinearity profile will be Gaussian as well.

Most domain engineering designs have looked at a PMF with a flat phase, as this is necessary for high-purity state generation~\cite{Graffitti2018}. 
However, here we demonstrate that if purity is not relevant, carefully engineering this phase further increases flexibility and can lead to drastic improvements, such as higher brightness (see section~\ref{sec:new_PMF}). 

Previous domain engineering algorithms, such as in~\cite{Tambasco2016} and~\cite{Graffitti2017}, mainly use different configurations of domains of equal width. 
Graffitti \textit{et al.}~\cite{Graffitti2017} additionally used stochastic annealing to modify the individual domain widths slightly and therefore further improve the result. 
However, these algorithms impose an unnecessary restriction, since modern photolithography can produce arbitrary domain widths in any configuration~\cite{Myers_96,Shur2015}. Therefore, the optimization process is ultimately only limited by the minimal feature size $l_{\text{min}}$, usually smaller than the coherence length $l_{\text{c}}$, and the lithographic resolution $l_{\text{res}}$, which is usually orders of magnitude smaller than $l_{\text{min}}$. 

\subsection{A new algorithm utilizing variable domain widths}\label{sec:new_algor}

As discussed above, the target field within the crystal $A_{\text{target}}$ is determined from the desired PMF. The field generated by the poling profile for a given phase mismatch $\Delta k$ is denoted as $A(z,\Delta k)$. At each step of the algorithm, where an additional domain of width $l$ is added, the new field is updated via
 \begin{equation}\label{eq:e-field_1}
    A_{\text{new}}(z+l,\Delta k) = A(z,\Delta k)+   \int_{z}^{z+l}g(\tilde{z}) \, e^{i\, \Delta k \, \tilde{z}}\,  d\tilde{z}.
\end{equation}
This field is compared to the target field by evaluating the cost \begin{equation}\label{eq:cost_1}
\epsilon =  | A_{\text{new}}-A_{\text{target}} |.
\end{equation} For this the algorithm uses the following procedure: it starts by setting the first domain with a width equal to the minimum feature size of the manufacturing
process $l_{\text{min}}$ with positive poling.

\textbf{\begin{figure}[hbt!]
	\centering
	\includegraphics[width=14cm]{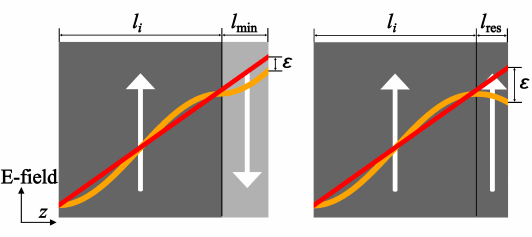}
	\caption{ \label{fig:algo_principle} Working principle of the domain engineering algorithm. As an example, one domain with the width of $l_i$ has already been computed. The white arrows indicate the poling direction. In the next step of the algorithm, either a new domain with opposite poling and the minimal domain width $l_{\text{min}}$ is added (left), or the current domain is extended by the domain resolution $l_{\text{res}}$ (right). For both these cases, the difference between the target field (red) and the computed realized field (orange) is calculated using equation~(\ref{eq:e-field_1}), and the option with lower deviation ($\epsilon_{j}$ from equation~(\ref{eq:cost_1})) is chosen. In this example, the new-domain-option (left) is closer to the target and is therefore chosen.}  
\end{figure}}

\begin {itemize}

\item Generate a candidate for \textit{swapped poling} (figure~\ref{fig:algo_principle} left) by introducing a new domain of the width $l_{\text{min}}$ and an opposing poling to the previous domain.

    \begin {itemize} 
     \item Calculate new field using equation~(\ref{eq:e-field_1}) and cost via equation~(\ref{eq:cost_1}).
    \end{itemize}

\item Generate a candidate for \textit{remain poling} (figure~\ref{fig:algo_principle} right) by extending the current domain by the photolithographic resolution $l_{\text{res}}$.

    \begin {itemize} 
     \item Calculate new field using equation~(\ref{eq:e-field_1}) and cost via equation~(\ref{eq:cost_1}).
    \end{itemize}

\item Find the candidate with lower cost and adapt the domain structure accordingly.
\item Repeat until the end of the crystal is reached. 

\end{itemize}

A first successful application of this algorithm was used to create hyperentanglement in polarization and frequency through joint spectrum shaping~\cite{faleo2026}.

\subsection{Algorithm comparison}\label{sec:algo_comp}
\begin{figure*}[htb!]
   \centering
    \includegraphics
    {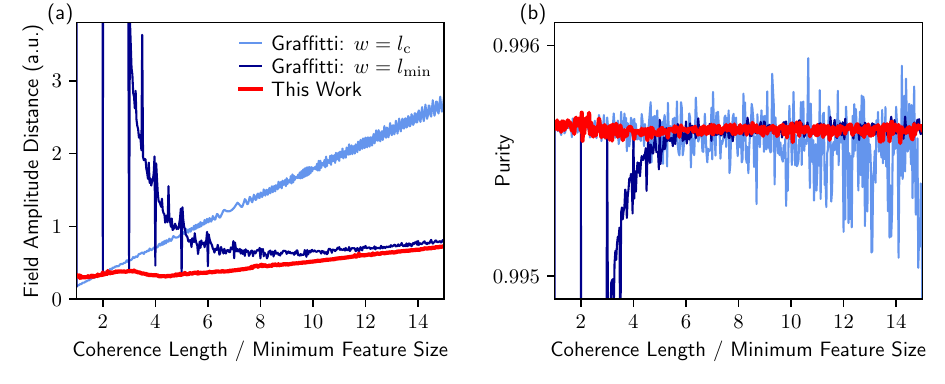}
\caption{\label{fig:algo_comp}Comparison of algorithms for Gaussian phase-matching in \SI{30}{mm}-long KTP for group velocity matched degenerate photon pairs at \SI{1582}{nm}: For the algorithm by Graffitti \textit{et al.}~\cite{Graffitti2017} two settings need to be considered: the \emph{coherence-length} setting:
$w=l_{\mathrm{c}}$, and the \emph{minimum-domain-width} setting: $w=l_{\text{min}}$. For the algorithm presented in this work, the parameters do not need to be adjusted.
(a) Distance between target and computed field as a function of minimum feature size. (b) Spectral purity from Schmidt decomposition~\cite{Zielnicki2018} as a function of minimum feature size.}
\end{figure*}

To benchmark the proposed algorithm, we compare its performance against the method of Graffitti \textit{et al.}~\cite{Graffitti2017} (without annealing) for real-valued Gaussian phase matching. We simulate the SPDC process in a \SI{30}{mm}-long KTP crystal and assume a pump wavelength of \SI{791}{nm} producing degenerate down-converted photons at \SI{1582}{nm}. This wavelength combination allows for group velocity matching, and therefore high spectral purity~\cite{Keller1997,Grice2001}. Since the fixed domain width, $w$, in the method of Graffitti \textit{et al.} must be selected manually, we consider two choices: (i) $w=l_{\mathrm{c}}=\SI{23}{\micro m}$, referred to as the \emph{coherence-length} setting; and (ii) $w=l_{\min}$, referred to as the \emph{minimum-domain-width} setting.
In figure~\ref{fig:algo_comp}(a), the computed field amplitude distance (mean absolute error) from the target field is shown for different minimum feature sizes.
Although for large and small minimum feature sizes, the resulting structure's performance is practically identical to that of~\cite{Graffitti2017} (matching the \emph{coherence-length} setting for $l_{\text{c}}\approx l_{\text{min}}$ and the \emph{minimum-domain-width} setting for $l_{\text{c}}\gg l_{\text{min}}$), we find improved performance for moderate minimum feature sizes on the order of $ l_{\text{c}}/l_{\text{min}}\approx 4$.
In this regime, we also find an increase in spectral purity as shown in figure~\ref{fig:algo_comp}(b). 
Generally, for this work, the purity is more stable over a wide range of manufacturing parameters. 
The underlying reason for this is that with the added flexibility, the algorithm is able to follow the target field more closely, as further explained in appendix~\ref{sec:app_algo_comp}. In conclusion, by allowing for more flexibility, the algorithm presented here maintains better performance across a wider range of coherence lengths without parameter tuning and removes the need to choose the right setting as in Graffiti \textit{et al}.

\section{An application of domain engineering for quantum sensing}\label{sec:bright_broad_source}
The example we use to demonstrate the design process and this new algorithm centers on a quantum light source for a novel optical coherence tomography (OCT) approach using time-energy entangled photon pairs and Franson interference~\cite{Franson1989,schlosser_OCT}. 
We aim for degenerate photons centered at \SI{809.4}{nm} and a Gaussian spectrum with a width of \SI{45}{\nano m} (full width at half maximum (FWHM)). In conventional OCT, this leads to an axial resolution of \SI{6.4}{\micro \meter}~\cite{OCT1991}. 

OCT using the Hong-Ou-Mandel effect~\cite{hong1987} (also termed Quantum-OCT) has already been proposed~\cite{Abouraddy2002} and demonstrated~\cite{Nasr2003}, providing a twofold resolution enhancement compared to its classical counterpart. 
Record axial resolution has been achieved by using a chirped quasi-phase-matched lithium tantalate device, where the poling period was linearly increased along the crystal~\cite{Okano2015}. 
This allows multiple wavelengths to phase-match at different positions within the crystal, thereby increasing the spectral bandwidth. 
Although a Quantum-OCT resolution of \SI{0.54}{\micro m} could be reached, the interference pattern exhibits strong undesired sidelobes, stemming from the non-Gaussian spectral shape, which in turn arises from the chirped quasi-phase-matching used. Most thus far explored sources for broadband photons have used such simple, strictly linear chirping of the poling period, even though there is great potential in incorporating truly aperiodic structures~\cite{Weiss2025}.
It would also be possible to obtain a large spectral bandwidth and, therefore, a high OCT resolution, using a very thin nonlinear crystal, which naturally allows for a larger phase mismatch~\cite{Katamadze2015}. 
However, such a small length of only a few hundred micrometers to a few millimeters leads to a very small interaction region between pump and down-converted fields, resulting in low total photon flux. 
By simply extending the length of the crystal, this flux increases, but the bandwidth also reduces, necessitating a trade-off between the two observables. 
In contrast, we present here a different approach that avoids this trade-off. 

\subsection{Improved quasi-phase-matching}\label{sec:new_PMF}
\begin{figure*}[htb!]
   \centering
    \includegraphics
    {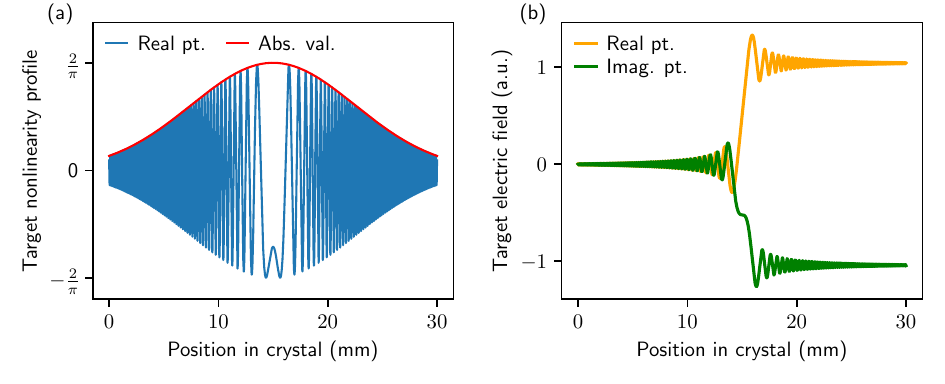}
 \caption{\label{fig:chi_eff_comp}(a) Analytically computed nonlinearity profile $g_{\text{target}}(z)$ inside the crystal of length $L=\SI{30}{mm}$, with $\sigma_{\chi}=L/4$ and $\sigma_{\phi}$ corresponding to a bandwidth of $\SI{45}{nm}$. (b) The resulting target field $A_{\text{target}}(z)$. Analytical expressions are presented in appendix~\ref{sec:app1}.}
\end{figure*}

Since the envisioned OCT-scheme does not require interference between independent photons, the source does not need to produce spectrally pure states, and we can imprint a quadratic phase on the PMF
\begin{equation}\label{eq:chirped_PMF}
\phi_{\text{target}} (\Delta k) =  \exp \left(- \frac{ \Delta k ^2 }{2 \,\sigma_{\phi}^2} \right) \cdot \exp \left( i \, a \, \Delta k ^2\right),
\end{equation}
where $\sigma_{\phi}$ is the width that eventually determines the spectral bandwidth of the down-converted photons (see equation~(\ref{eq:lc_def}) and equation~(\ref{eq:sigma_phi}), with $k=2\pi n/\lambda$), and $a$ is the quadratic phase coefficient, used to increase the width of the nonlinearity profile, thus increasing brightness. This is ultimately limited by the available crystal length.
From this desired PMF, the target nonlinearity profile can be computed (equation~(\ref{eq:target_nonlinear})) and, in turn, using equation~(\ref{eq:target_field}), the target field can be calculated for each position within the crystal. 
Analytical expressions are presented in appendix~\ref{sec:app1}, and they are also visualized in figure~\ref{fig:chi_eff_comp}. 
From equation~(\ref{eq:chirped_PMF}), the quadratic phase coefficient is
\begin{equation}\label{eq:phase_coeff}
a = \frac{\sqrt{ \,\sigma_{\chi}^2\, \sigma_{\phi}^2 - 1}}{2\,\sigma_{\phi}^{2}},
\end{equation}
with $\sigma_{\chi}$ being the width (standard deviation) of the absolute value of the nonlinearity profile within the crystal. 
As suggested by Tambasco \textit{et al.}~\cite{Tambasco2016}, a reasonable value is $\sigma_{\chi}=L/4$, with $L$ denoting the length of the crystal, which is a compromise between accuracy to the desired shape and brightness. 
A larger $\sigma_{\chi}$ would increase brightness at the cost of inducing oscillations in the PMF as a result of the truncation of the nonlinearity profile at the crystal boundaries.

Note that, even though the PMF has a quadratic phase, which equates to a linear chirp, this design is fundamentally distinct from linearly increasing the poling period along the crystal, which is called chirped quasi-phase-matching~\cite{Maine1988,Carrasco2004,Charbonneau-Lefort08}.
The main difference is that we obtain a highly nontrivial poling pattern (see appendix~\ref{sec:domain_analysis}), since this approach utilizes more degrees of freedom, increasing the flexibility. 
This allows the generation of customizable spectral shapes. 
With this design, substantial increases in spectral bandwidth lead to only minor reductions in overall brightness, underscoring that this undesirable trade-off has been largely mitigated, as shown in appendix~\ref{sec:design_simulations} in more detail.

In our application example, we choose collinear type-II SPDC in KTP and deploy the algorithm described in section~\ref{sec:new_algor} to find a domain structure matching the target field. 
The resulting electric field matches the target within the thickness of the lines used to visualize them in figure~\ref{fig:chi_eff_comp}(b). We numerically verify the PMF resulting from this domain structure for the entire two-dimensional joint spectral state and show the result in figure~\ref{fig:signal_idler_spectrum_with_simulation}(a), together with a visualization of the amplitude of a narrowband pump laser. The product of these two functions gives rise to the joint spectral amplitude (JSA) of this SPDC output state.

\subsection{Experimental results}\label{sec:experimental_results}
The crystal length is constrained by the lithography fabrication process, and is therefore fixed at a current achievable maximum of \SI{30}{mm}. The crystal was commercially fabricated by \textit{SLF Quantum} for a signal-idler wavelength-degeneracy temperature of \SI{50}{\degreeCelsius}. It is placed inside a temperature-controlled oven and put into an optical setup to test its performance. The setup consists of a fiber-coupled continuous wave pump laser at $\lambda_{\text{p}}=\SI{404.7}{nm}$, which is polarization filtered using a polarizing beam splitter (PBS) to ensure only the relevant horizontally polarized light reaches the crystal. The pump is focused into the center of the crystal to achieve a waist radius of $\approx \SI{15}{\micro m}$, and the signal, idler, and pump modes are collimated again after the crystal. A dichroic mirror and a hard-coated longpass filter are used to filter out the remaining pump light.
 
The orthogonally polarized down-converted photons are deterministically split using another PBS before being coupled into separate single-mode fibers for detection. Using superconducting nanowire single-photon detectors (quantum efficiency $\SI{88(3)}{\%}$), we were able to measure single photon count rates of up to $R_{\text{signal}} \approx R_{\text{idler}} \approx  \SI{200}{kHz}$ per $\SI{}{mW}$ of pump power and a pair rate of $R_{\text{pair}}\approx \SI{40}{kHz}$ per $\SI{}{mW}$, resulting in a raw, uncorrected heralding efficiency of $\eta =R_{\text{pair}} \,  /\sqrt{R_{\text{signal}}\, R_{\text{idler}} }  \approx \SI{20}{\%}$.

\begin{figure*}[htb!]
   \centering
    \includegraphics
    {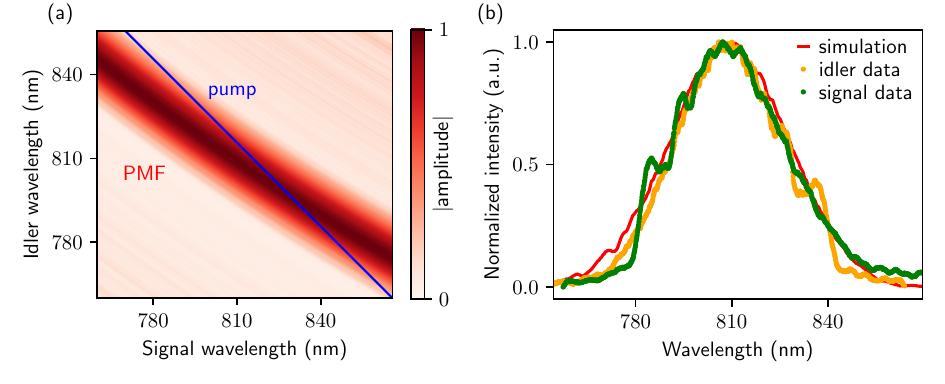}
 \caption{\label{fig:signal_idler_spectrum_with_simulation}(a) Simulated PMF and pump amplitude. This two-dimensional visualization of the PMF shows the amplitude of a given wavelength pair created by the crystal. The blue line labeled "pump" represents all wavelength pairs that adhere to conservation of energy for a narrowband pump at \SI{404.7}{nm}. (b) Experimentally measured signal and idler spectra at a crystal temperature of \SI{70}{\degreeCelsius}, in comparison to a simulated spectrum using a plane wave propagated through the crystal. To reduce high-frequency noise, a moving average with a window of \SI{2}{nm} (20 data points) was used. 
 }
\end{figure*}
The individual spectral intensities were measured using a grating spectrometer, and the results are shown in figure~\ref{fig:signal_idler_spectrum_with_simulation}(b) for the measured degeneracy temperature of $\approx\SI{70}{\degreeCelsius}$. A comparison to a simulated spectrum (which is approximately identical for signal and idler), for which plane waves are numerically propagated through the crystal, shows good agreement. The largest deviation between simulation and measurement can be seen for the signal photon's spectrum between \SI{780}{nm} and \SI{790}{nm}. A mirror image of this deviation is present in the idler spectrum at about $\SI{840}{nm}$, too. 
This suggests some minor imperfections in the PMF itself, probably arising from inaccuracies in the fabricated nonlinearity profile. 
Note that there is a small discrepancy between the desired spectral bandwidth of \SI{45}{nm} and a fit to the result of the simulation with \SI{43.5}{nm}, stemming from discretization errors when approximating the smooth target nonlinearity with a real and finite binary step function. This highlights the need to verify the generated poling structure and to possibly compensate accordingly. After manufacturing the crystal, we found a slightly modified design approach addressing this issue, and we present simulations in section~\ref{sec:demod}.

Even the here implemented source, as designed with the basic version of our algorithm, roughly produces the same pair rate as a similar setup containing a \SI{10}{mm} long periodically poled KTP crystal with a poling period of \SI{9.9}{\micro m}. However, we achieve over 50 times larger spectral bandwidth.

\section{Step-by-step guide}\label{sec:setpbystep}
Here we provide a tutorial on how to approach the domain engineering process in several steps. 

\begin{enumerate}
   
    \item Define desired properties of the photon pair:\\
    An important consideration is whether the state needs to be spectrally pure. This is the case if independent photons from such states interfere with one another~\cite{Zielnicki2018}. If no such interference is desired, one can relax this purity constraint.


    \item Note that, as evident from equation~(\ref{eq:target_field}), domain engineering can only alter the PMF along the gradient of 
    \begin{equation}\label{eq:delta_k_def}
           \Delta k (\lambda_\text{s}\, , \lambda_\text{i} ) = 2\pi \left(  \frac{n_\text{p}
           (\lambda_\text{p})}{\lambda_\text{p} }- \frac{n_\text{s}
           (\lambda_\text{s})}{\lambda_\text{s} }- \frac{n_\text{i}(\lambda_\text{i})}{\lambda_\text{i} } \right),
    \end{equation}
    with $\lambda_\text{p}^{-1}=\lambda_\text{s}^{-1} +\lambda_\text{i}^{-1}$.
    Precise knowledge of the Sellmeier coefficients for the given material is therefore necessary~\cite{Fradkin1999,konig2004,Emanueli2003}. By examining this $\Delta k$-landscape in the wavelength region of interest, one needs to consider whether domain engineered quasi-phase-matching provides an advantage over birefringent phase-matching. Furthermore, since the various types of SPDC require different methods for separating signal and idler photons, a suitable configuration needs to be selected.


    \item Find the PMF in $\Delta k$-space:\\
	Using equation~(\ref{eq:lc_def}) and equation~(\ref{eq:bandwidth_coherence}), one can express the desired central wavelength and bandwidth in terms of coherence lengths of the nonlinear process. This
    allows a definition of a PMF, such as the chirped Gaussian of equation~(\ref{eq:chirped_PMF}) chosen here as an example. \\
    Small discrepancies between the desired and fabricated poling structure can change the down-converted photon properties. Most notably, the temperature at which signal and idler wavelengths are degenerate is hard to predetermine precisely.  For this reason, it is advisable to set this degeneracy temperature several tens of degrees Celsius above room temperature. This allows easy temperature control using resistive heaters in an oven. 

        
    \item Fourier transform the desired PMF into an effective nonlinearity profile $g_{\text{target}}(z)$ and normalize it so that its maximum (here located in the center of the crystal) coincides with the maximum achievable by periodic poling for the central wavelengths, i.e., $g_{\text{target}}(L/2) = 2/\pi$~\cite{Tambasco2016}.
         
    
    \item The target field $A_{\text{target}}(z)$ can now be calculated analytically using equation~(\ref{eq:target_field}).
    Note that it is sufficient to perform these calculations for only one $\Delta k$, significantly reducing computational costs. 

   
    \item Specify manufacturing parameters (crystal length, minimal domain width, and domain resolution) and run the tracking algorithm. Export the start and endpoints, as well as the poling orientation, for all the domains.  


    \item Validate the domain structure using a numerical simulation that propagates the entire two-dimensional joint field (as a function of wavelengths $\lambda_s$, $\lambda_i$; for narrow-band pumps the variation of a single wavelength is sufficient) through the crystal using equation~(\ref{eq:e-field_1}). Note that this task is much more computationally expensive than running the algorithm itself. However, it is highly parallelizable~\cite{Masoudi1996}. 
    Note that, depending on the requirements, a simple plane wave propagation is sufficient, or, in the case of tight focusing, the spatial mode profiles of the pump, signal, and idler fields need to be considered as well~\cite{Kolenderski2009, Bennink2010}. The JSA obtained in this way can now be analysed. For example, one can estimate the spectral purity via Schmidt decomposition~\cite{Zielnicki2018} or project the JSA onto signal and idler axes, respectively, to obtain the marginal spectral distributions. These can be compared to the desired spectra from step 1.

\end{enumerate}

\section{Further improvements: spectral shape optimization}\label{sec:improvments}
Here we propose additional improvements to the algorithm.
These modifications were only found after the crystal was manufactured and tested, which is why no experimental verification was performed, and we show here only simulation results.
As discussed above, the greedy optimization of domain engineering algorithms tracks the target field by matching its complex value at a single phase mismatch $\Delta k_0$.  
While this is sufficient to approximate the target PMF at a specific phase mismatch, $\phi_{\text{target}}(\Delta k_0)$, it does not guarantee that the PMF is approximated well for different values of $\Delta k$.
Considering $\Delta k = \Delta k_0 + q$, the deviation between the realized and ideal target PMF, $\delta \phi(\Delta k) = \phi(\Delta k) - \phi_{\text{target}}(\Delta k)$, at detuning $q$ is (see derivation in appendix~\ref{app:target_deviation})
\begin{equation}\label{eq:error_q}
    \delta \phi(q) = \delta A(L)e^{iqL} -iq \int_0^L \delta A(z) e^{i q z} \mathrm{d}z \, +\epsilon_{\text{L}}(q),
\end{equation}
where $\delta A(z) = A(z) -A_{\text{target}}(z)$ is the tracking error at each position $z$ and $\epsilon_{\text{L}}$ is the difference between the target PMF evaluated over the finite crystal length and the ideal target PMF. This difference is inherent to any realized poling structure, and its magnitude depends on $\sigma_{\chi}/L$; i.e., the more gradually the envelope of the nonlinearity approaches zero, the lower $\epsilon_{\text{L}}$.
This shows that, unless $\delta A(z)=0 \; \forall z \in [0, L]$, which is generally not possible with only a binary poling, the tracking error is weighted by $q$, potentially introducing distortions at large enough $q$. 

\subsection{Demodulation}\label{sec:demod}

To give a first demonstration of this effect, we analyze again the design in section~\ref{sec:bright_broad_source} with a different approach.
So far, the algorithm has tried to minimize the distance between the realized field 
\begin{equation} \label{eq:running_field}
    A(z) =  \int_{0}^{z}g(\tilde{z}) \, e^{i\, \Delta k \, \tilde{z}}\,  d\tilde{z}
\end{equation}
and $A_{\text{target}}(z)$, as defined in equation~(\ref{eq:target_field}).
Here, to improve upon this, we decompose the target nonlinearity profile into its amplitude and phase $g_{\text{target}}(z)=\left| g_{\text{target}}(z) \right| \cdot e^{i\theta(z)}$, and we define new functions 
\begin{equation}
    B(z) =  \int_{0}^{z}g(\tilde{z}) \, e^{i[ \Delta k \, \tilde{z} -\theta(\tilde{z})]}\,  d\tilde{z}
    \quad
    \text{ and}
    \quad
    B_{\text{target}}(z) =  \int_{0}^{z}\left|g_{\text{target}}(\tilde{z})\right| \, \,  d\tilde{z},
\end{equation}
which corresponds to a change of reference frame.
This transformation acts on the integrand of the field $B(z)$, such that the algorithm generally minimizes a different quantity while leaving the target PMF unchanged.

\begin{figure*}[htb!]
   \centering
    \includegraphics
    {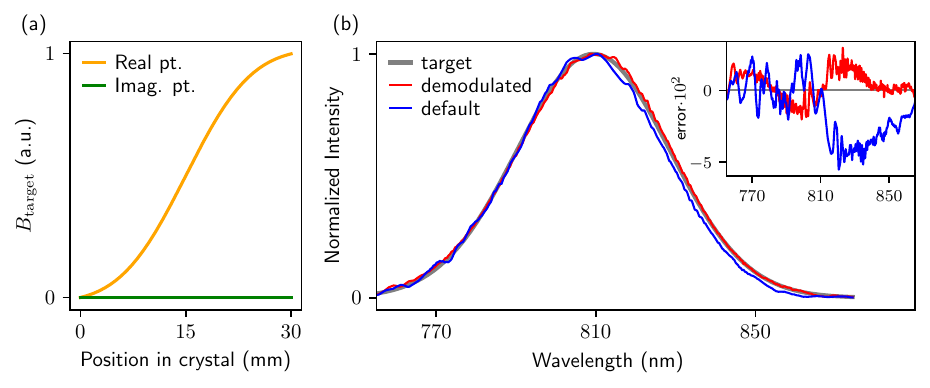}
 \caption{\label{fig:demod}(a) Real and imaginary part of the normalized demodulated target field $B_{\text{target}}(z)$. (b) Comparison of the computed signal intensities from the original design (default) and the demodulation approach to the target function. The inset shows the difference in intensities to the target. To reduce high-frequency noise, a moving average with a window of \SI{2}{nm} (20 data points) was used.}
\end{figure*}

In contrast to the complex oscillating target field $A_{\text{target}}(z)$ of figure~\ref{fig:chi_eff_comp}(b), the new target $B_{\text{target}}(z)$ is a smooth real function, as shown in figure~\ref{fig:demod}(a).
Thus, minimizing the distance $|B - B_{\text{target}}|$ and therefore $\delta\phi$, analogously to the procedure outlined in section~\ref{sec:new_algor}, mitigates the effect of the inevitable discretization errors on the realized PMF at $\Delta k \neq \Delta k_0$. Therefore, the target PMF can be better approximated over the entire range of frequencies (see details in appendix~\ref{sec:app_demod}).

This is demonstrated in figure~\ref{fig:demod}(b), which shows the computed signal spectrum at signal-idler wavelength degeneracy for both approaches.
The demodulated case more closely approximates the desired function by reducing the spectral distortions away from the phase-matched wavelength. 
The Gaussian fit to the simulated spectrum of the demodulated approach yields a FWHM of \SI{45}{nm}, in precise agreement with the initially desired value. 
Note that the simulated degeneracy temperature shifts slightly between the two approaches, from \SI{50}{\degreeCelsius} in the original to \SI{65}{\degreeCelsius} in the demodulated case.
This modification improves only the spectral shape of the down-converted photons, leaving the overall flux approximately unchanged. 
Furthermore, for PMFs with a flat phase, i.e. $ \theta (z)= \text{const.}$ (as used for the comparisons in section~\ref{sec:algo_comp}), this approach defaults back to the original algorithm.

\subsection{Structure Refinement}\label{sec:refinement}

Motivated by the observations of equation~(\ref{eq:error_q}) and the result of the previous section, we introduce here a new and efficient refinement algorithm that deterministically and locally optimizes the initial poling structure produced by the greedy optimizer. This simultaneously reduces the complex PMF error across a set of $\Delta k$ values while respecting the fabrication constraints.
The algorithm varies the position of each domain wall resulting from the initial optimisation in sequence from $z=0$ to $z=L$, evaluating the effect of positive and negative displacements, applied in integer steps of the minimal resolution $l_{\text{res}}$, on the complex PMF error,
\begin{equation} \label{eq:objective}
    J = \sum_{\Delta k} \xi(\Delta k) \left| \phi(\Delta k) - \phi_{\text{target}}(\Delta k) \right|^2,
\end{equation}
where $\xi(\Delta k)$ is a weighting function that prioritizes specific $\Delta k$ values (see also appendix~\ref{app:weighting_function}).
Only displacements that preserve a minimum width $w \geq l_{\text{min}}$ of neighbouring domains are considered, and the one giving the largest reduction of $J$ is accepted when the reduction exceeds a minimum threshold.
Otherwise, the wall is left unchanged.
In addition to these displacements, the algorithm also considers the removal of a pair of adjacent walls: three domains of alternating sign collapse into a single domain with same total width, provided this reduces $J$.
No walls are inserted, so the number of domains can only decrease, and the total crystal length is preserved.
This whole procedure is performed iteratively, until no reduction of $J$ above the minimum threshold is found. The refinement algorithm is also described step-by-step in appendix~\ref{app:refinement_procedure}.

This new refinement procedure has been applied in simulation to the designs presented above (default and demodulated) and both cases lead to an improvement of the final PMF (see more details and comparisons in appendix~\ref{app:refinement_results}). The result of the refinement procedure remains directly manufacturable, as every displacement is an integer multiple of $l_{\text{res}}$ and $l_{\text{min}}$ is enforced, so no further discretisation step is needed.
The algorithm is inexpensive (see appendix~\ref{app:efficiency}), accepts only reductions of $J$ starting from an input structure, does not involve the randomness of simulated annealing~\cite{Graffitti2017}, provides the same output at every run, and converges in tens to hundreds of iterations.
As noted before, it does not provide a global optimizer and must be thought of as a cheap optimization of tracked structures obtained with domain engineering algorithms that can result in significant improvements.

\section*{Conclusion and outlook}\label{sec:conclusion}
The deterministic domain engineering algorithm presented here utilizes the full manufacturing potential of nonlinear crystals and, thereby, achieves smaller deviations between the target and generated down-converted fields compared to previous algorithms.
We provide a detailed practical guide on how to use this algorithm, including a step-by-step tutorial and the source code written in Python. This guide is aimed at researchers requiring nonlinear optical sources, and it helps in all phases of the design process, from choosing sensible desired photon properties to final verification. 

An example of such a design process is given for a \SI{30}{mm}-long KTP crystal used as a bright and broadband photon pair source for quantum sensing. By understanding the experimental requirements, we create a design that utilizes the maximum crystal length currently manufacturable, leading to strong nonlinear interaction and therefore a high pair rate of \SI{40}{kHz} per mW for degenerate photon pairs centered at \SI{809}{nm}, with quasi-Gaussian spectral shape and a FWHM-bandwidth of \SI{43.5}{nm}. The commercially fabricated domain structure has been experimentally tested and shows good agreement with the expected results.  

Further improvements to the algorithm are discussed and show potential for the realization of complex yet precise PMFs. We hope that this new algorithm, in tandem with this guide, further encourages the use of domain engineering in quantum light sources. This can allow a great variety of technologies to benefit from this mostly untapped potential.

\section*{Source code/data availability}
\noindent
An implementation of the presented algorithm, written in Python, is available at: \url{https://git.uibk.ac.at/c7041192/apyriodic-poling-public} 

\noindent
The data and code that support the findings of this study are openly available in reference~\cite{schlosser_2026_zenodo}.

\section*{Funding}
\noindent
This research was funded in part by the Austrian Science Fund (FWF), projects 10.55776/F71 and 10.55776/PIN6357923 and the MEANTIME 1669 Prototype Project funded through the 1669 Förderverein.

\section*{References}\label{sec:references}

\bibliography{references.bib}

\appendix

\section{Algorithm's enhanced flexibility}\label{sec:app_algo_comp}
\begin{figure*}[htb!]
   \centering
	\begin{tikzpicture}
        \node (alabel) {(a)};
	    \node[anchor=north west] (pica) at (alabel.north east) {\includegraphics[width=.42\textwidth]{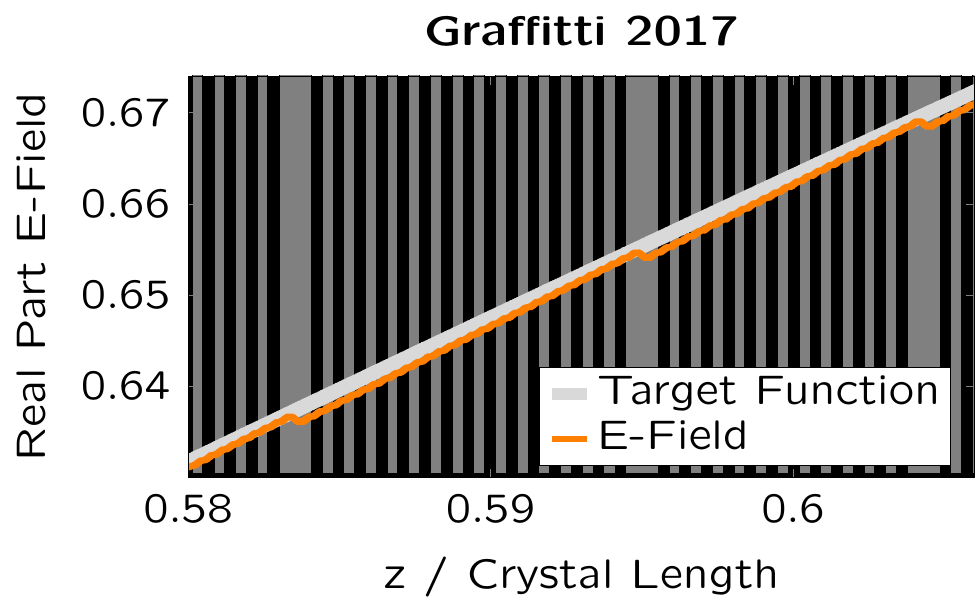}};
        
        \node[anchor=north west] (blabel) at (pica.north east) {(b)};
        \node[anchor=north west] (picb) at (blabel.north east)
        {\includegraphics[width=.42\textwidth]{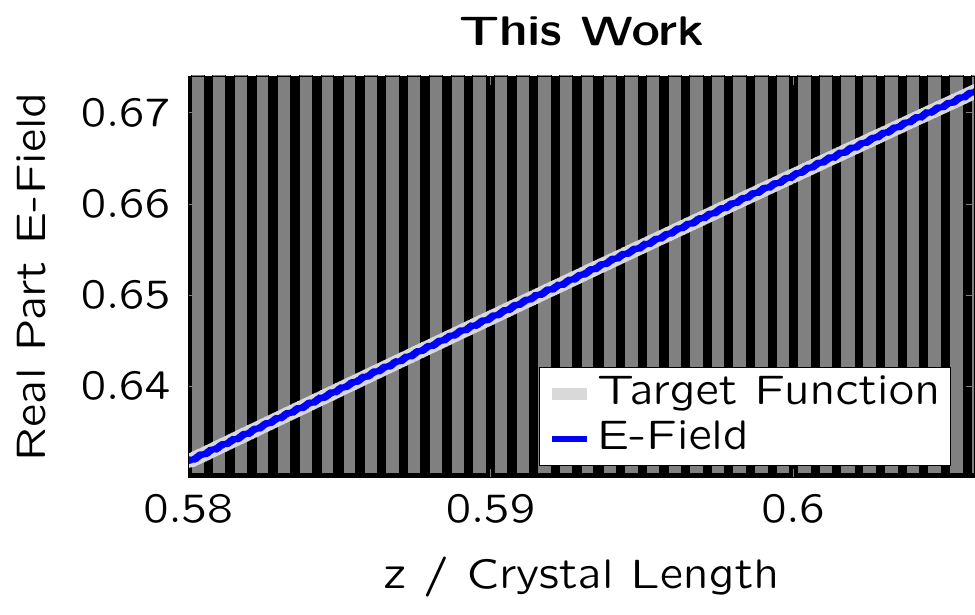}};
        
        \node[anchor=north east] (clabel) at (pica.south west) {(c)};
        \node[anchor=north west] (picc) at (clabel.north east)
        {\includegraphics[width=.42\textwidth]{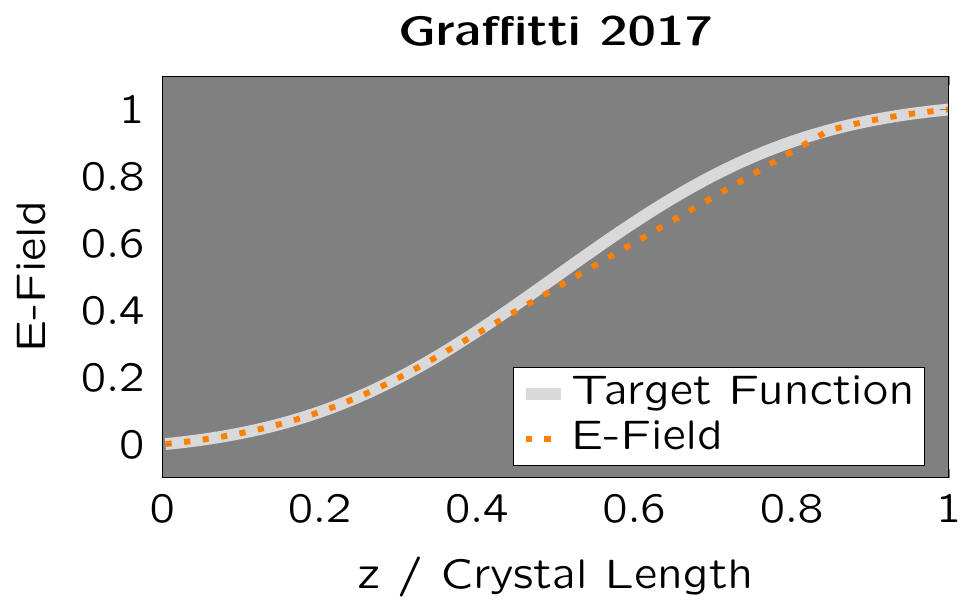}};

        \node[anchor=north east] (dlabel) at (picb.south west) {(d)};
        \node[anchor=north west] (picd) at (dlabel.north east)
        {\includegraphics[width=.42\textwidth]{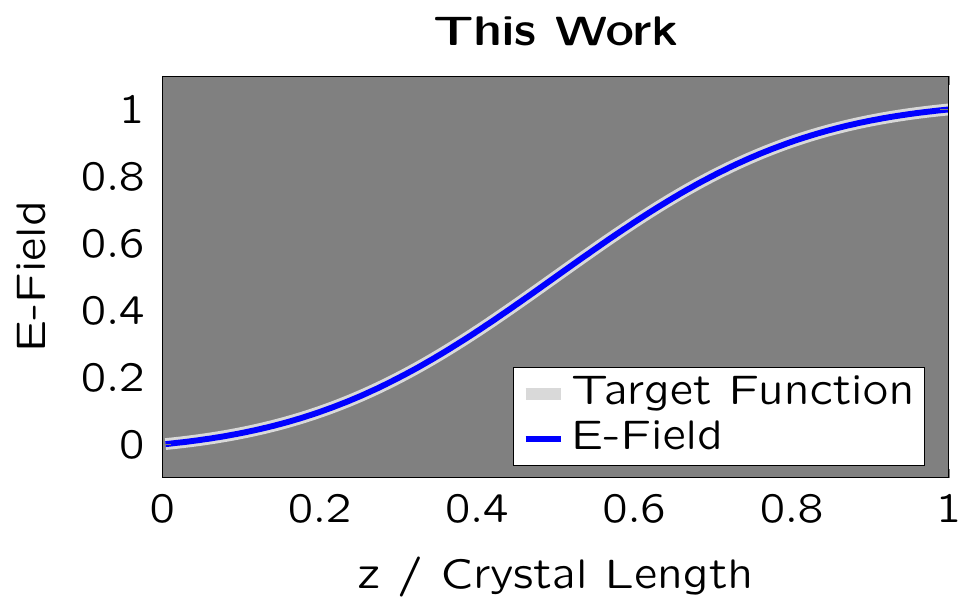}};

	\end{tikzpicture}
 \caption{\label{fig:algo_comp_appendix}Visual explanation of the enhanced performance in comparison to the algorithm by Graffitti \textit{et al.}~\cite{Graffitti2017}.  (a-b) A large discretization when the domain width equals the coherence length leads to artifacts. (c-d) Small domain widths ($w = l_{\text{min}}$) fail to produce strong effective nonlinearity for coherence lengths that are not divisible by the domain widths. For both regimes, our ability to set each domain width individually leads to a smaller discrepancy from the target field. For~\cite{Graffitti2017}, one needs to set one domain width for the entire crystal, leading to less flexibility.}
\end{figure*}

When compared to Graffitti \textit{et al.}~\cite{Graffitti2017}, our work achieves better performance because the variable domain widths add flexibility (see section~\ref{sec:algo_comp}). This manifests in a closer match between the generated field to the target field, as illustrated in figure~\ref{fig:algo_comp_appendix}. For the \emph{coherence-length} setting of~\cite{Graffitti2017}, i.e., $w = l_{\text{c}}$ (figure~\ref{fig:algo_comp_appendix}(a)), the algorithm fails to closely follow the target field when its slope is not at a maximum. For the used example of Gaussian phase-matching, this maximum in slope is only reached at the center of the crystal, while everywhere else it is smaller. Since our algorithm allows for domains with arbitrary widths, these smaller slopes can be followed more closely, as shown in figure~\ref{fig:algo_comp_appendix}(b). 
When considering small domain widths close to the resolution limit ($w = l_{\text{min}}$), the algorithm of~\cite{Graffitti2017} fails to produce a strong effective nonlinearity for coherence lengths which are not divisible by $l_{\text{min}}$, see figure~\ref{fig:algo_comp_appendix}(c). The reason for this is that the strongest nonlinear conversion takes place if the domain width equals the coherence length. 
However, $l_{c}$ will generally not be divisible by $l_{\text{min}}$, such that after one coherence length there will be one domain that causes partially destructive interference, leading to an overall decrease in field amplitude. Again, our work reduces this discrepancy, as shown in figure~\ref{fig:algo_comp_appendix}(d). Therefore, the distance to the target field can be decreased without parameter tuning, using the additional flexibility of the new algorithm.

\section{Analytical forms of expressions}\label{sec:app1}
Note that the PMF is a function of $\Delta k$ and the computations are conveniently expressed in terms of the central coherence length $l_{\text{c}}$ and 
\begin{equation}\label{eq:bandwidth_coherence}
    \delta l_{\text{c}} = \left| \frac{\pi}{ \Delta k (\lambda_\text{s-}  , \lambda_\text{i-} ) }-\frac{\pi}{ \Delta k (\lambda_\text{s+}  , \lambda_\text{i+} ) } \right|,
\end{equation}
the bandwidth of the nonlinear process, with $\lambda_{\text{s}\pm}=\lambda_\text{s}\pm \Delta \lambda/2$ and $1/\lambda_{\text{i}\pm}=1/\lambda_\text{s}+1/\lambda_\text{i}-1/\lambda_{\text{s}\pm}$ for a spectral bandwidth of $\Delta \lambda$ and central wavelengths $\lambda_\text{s,i}$. $\Delta k$ is defined in equation~(\ref{eq:delta_k_def}). We therefore use
\begin{equation}\label{eq:sigma_phi}
    \sigma_{\phi}  = \frac{\pi}{2\sqrt{2\ln(2)}} \cdot\frac{ \delta l_{\text{c}}}{l_{\text{c}}^2 - \frac{\delta l_{\text{c}}^2}{4}}.
\end{equation}
The Gaussian PMF with a quadratic phase (example given above) is 
\begin{equation}\label{eq:chirped_PMF_app}
\phi_{\text{target}} (\Delta k) =  \exp \left(- \frac{ \Delta k ^2 }{2 \,\sigma_{\phi}^2} \right) \cdot \exp \left( i \, a \, \Delta k ^2\right),
\end{equation}
where the quadratic phase broadens the nonlinearity profile, increasing photon pair flux. Via a Fourier transform, we find the effective nonlinearity profile
\begin{equation}\label{eq:non_lin_analytical_chriped}
g_{\text{target}}(z)=\frac{2}{\pi} \exp \left(-\frac{i \pi ^2 \delta l_{\text{c}}^2 z^2}{4 \pi ^2 a \delta l_{\text{c}}^2+i \ln (2) \left(\delta l_{\text{c}}^2-4 l_{\text{c}}^2\right)^2}\right),
\end{equation}
which is shown in figure~\ref{fig:chi_eff_comp}(a).
This leads to a target field of the form
\begin{eqnarray}
 A_{\text{target}}(z) = \frac{\sqrt{\ln (2) \left(\delta l_{\text{c}} ^2-4 l_{\text{c}}^2\right)^2-4 i \pi ^2 a \delta l_{\text{c}} ^2} }{\pi ^{3/2} \delta l_{\text{c}} } \cdot \quad  \quad \quad \quad  \quad \quad \quad \quad \nonumber \\
 \left[\text{erf}\left(\frac{\pi  \delta l_{\text{c}}  L}{2 \sqrt{\ln (2) \left(\delta l_{\text{c}} ^2-4 l_{\text{c}}^2\right)^2-4 i \pi ^2 a \delta l_{\text{c}} ^2}}\right)+    
 \text{erf}\left(\frac{\pi  \delta l_{\text{c}}  \left(z-\frac{L}{2}\right)}{\sqrt{\ln (2) \left(\delta l_{\text{c}} ^2-4 l_{\text{c}}^2\right)^2-4 i \pi ^2 a \delta l_{\text{c}} ^2}}\right)\right]
\end{eqnarray}
which is shown in figure~\ref{fig:chi_eff_comp}(b).

\section{Domain structure analysis}\label{sec:domain_analysis}
\begin{figure*}[htb!]
   \centering
    \includegraphics
    {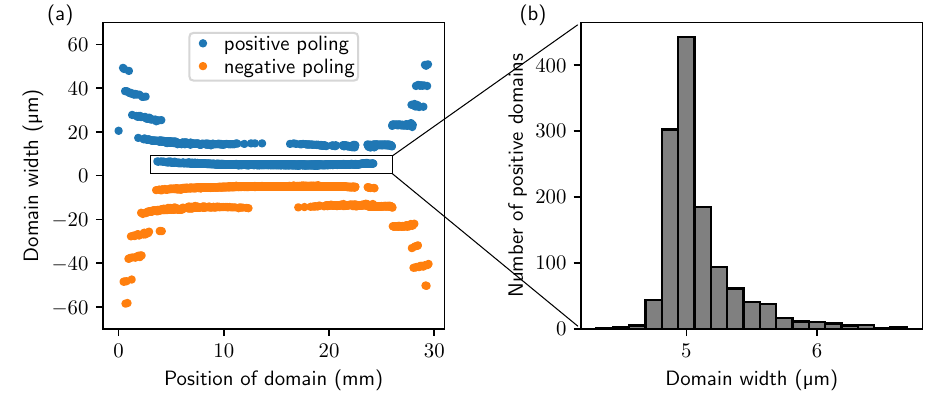}
    \caption{\label{fig:domain_analysis}(a) Visualization of the domain widths in the used design. For illustrative purposes, the width of each negatively poled domain is sign-inverted. (b) Histogram of positive domain widths close to the coherence length. The bin size is set to the domain resolution of \SI{125}{nm}}
\end{figure*}

For our requirements mentioned in section~\ref{sec:bright_broad_source}, the coherence length for the  central wavelengths correspond to $l_{\text{c}}=\SI{4.95}{\micro m}$, while the bandwidth corresponds to $\delta l_{\text{c}} = \SI{0.5}{\micro m}$. We use $a=\SI{134625}{\micro m ^2}$ for $L=\SI{30}{mm}$. The minimal domain size was chosen to be \SI{4}{\micro m} and the resolution \SI{125}{nm}. We obtain a structure with 3481 domains. The smallest domain size we obtain is \SI{4.375}{\micro m}. The different widths of the domains produced by the algorithm are shown in figure~\ref{fig:domain_analysis}(a). Note that one domain with a width of \SI{400}{\micro m} at the beginning, and one of \SI{500}{\micro m} at the end, are omitted for better visibility of the remaining domains. We see clear accumulations of domain widths at $\approx \SI{5}{\micro m}$ towards the center of the crystal, and accumulations of multiples of this value more towards the edges. This is expected, as it corresponds to the coherence length of this nonlinear process for the central wavelengths. A histogram of the domain widths around the coherence length is shown in figure~\ref{fig:domain_analysis}(b), where the bin size was set to the domain resolution. We observe that, although the largest bin of domains is the one with a width of exactly \SI{5}{\micro m}, a significant portion of domains are longer or shorter. In fact, \SI{65}{\%} of these domains have a width differing from \SI{5}{\micro m}. This shows that, when allowing for an arbitrary width for each domain individually, one achieves closer agreement between the target and the realized down-converted field.  

\section{\label{sec:design_simulations}Design parameter variations}
To further discuss the design introduced in section~\ref{sec:new_PMF}, we simulate a variation in the desired spectral bandwidth and calculate the resulting total photon pair flux, i.e., the integral over the entire joint spectral intensity, in the plane-wave limit, i.e., in the absence of focusing, for a \SI{30}{mm} long crystal. 
We find that the flux only varies slowly with bandwidth: even a large increase in bandwidth (from \SI{43.5}{nm} to \SI{81}{nm}) only reduces the flux by \SI{6.7}{\%}. Likewise, a reduction in bandwidth to \SI{22}{nm} leads to only a small increase in flux of $\SI{2.1}{\%}$.
A variation of the crystal length (for a fixed spectral bandwidth) shows a linear dependence of the total photon pair flux.

\section{Details Improvements}

\subsection{Target Deviation}\label{app:target_deviation}
We derive here equation~(\ref{eq:error_q}).
Let us consider the realized field in equation~(\ref{eq:running_field}), and the corresponding PMF,
\begin{equation}
    \phi(\Delta k) = \int_0^L g(z) e^{i \Delta k z} \mathrm{d}z.
\end{equation}
Introducing the detuning $q=\Delta k - \Delta k_0$, the PMF can be written as
\begin{equation}
    \phi(q) = \int_0^L g(z) e^{i \Delta k_0 z}e^{i q z} \mathrm{d}z = \int_0^L \frac{dA(z) }{dz} e^{i q z} \mathrm{d}z,
\end{equation}
where the second equality comes from the definition of the field in equation~(\ref{eq:running_field}). Integration by parts and the fact that $A(0)=0$ yield
\begin{equation}
    \phi(q) = A(L)e^{i q L} -iq\int_0^L A(z) e^{i q z} \mathrm{d}z,
\end{equation}
which reduces to the expected expression for $q=0$, i.e., $\phi(q=0)=A(L)=\phi(\Delta k_0)$.
Since we never used any specific property of $g(z)$, the same analysis holds true for the target PMF evaluated over the finite crystal length $\phi_{\text{target,L}}(q)=\int_{0}^{L}g_{\text{target}}(z)e^{iqz+i\Delta k_0z}dz$, and the deviation $\delta \phi_{\text{L}}(q) = \phi(q) -\phi_{\text{target,L}}(q)$ between the PMFs at $q$ results in
\begin{equation}\label{eq:delta_phi}
    \delta \phi_{\text{L}}(q) = \delta A(L)e^{iqL} -iq \int_0^L \delta A(z) e^{i q z} \mathrm{d}z.
\end{equation}

The deviation to the ideal target PMF can be obtained by adding $\phi_{\text{target,L}}(q)$ and subtracting $\phi_{\text{target}}(q)$, leading to equation~(\ref{eq:error_q}) with $\epsilon_{\text{L}}(q)=\phi_{\text{target,L}}(q) - \phi_{\text{target}}(q)$.

\subsection{Details Demodulation}\label{sec:app_demod}

The PMF defined in equation~(\ref{eq:chirped_PMF}) has a chirp, which leads to a phase in the nonlinearity profile.
This phase is reproduced through a local poling period that varies along the crystal, causing different down-converted frequencies to be generated at different positions within the crystal.
In fact, since the algorithm ensures $g(z)\,e^{i\Delta k_0 z}\approx g_{\text{target}}(z)=|g_{\text{target}}(z)|\,e^{i\theta(z)}$, the binary poling must contain a spatial component $g(z)\propto e^{-i[\Delta k_0 z - \theta(z)]}$ with local phase mismatch $\frac{d}{dz}[\Delta k_0 z-\theta(z)]=\Delta k_0-\theta'(z)$,  which equals the phase mismatch $\Delta k = \Delta k_0 + q$ when $\theta'(z)=-q$.
The down-converted frequency related to $\Delta k$ is, thus, generated only in a region of the crystal where this condition is approximately fulfilled, i.e., $\theta'(z) \approx -q$.
In the default frame, the deviation of the PMF is given by equation~(\ref{eq:delta_phi}), which depends on $q$, leading to larger errors at large $q$.
In the demodulated frame, however, the same integration by parts yields
\begin{equation}
    \delta \phi_{\text{L}}(q) = [B(L)-B_{\text{target}}(L)] e^{i[\theta(L)+qL]} -i\int_{0}^{L}(\theta'(z)+q)[B(z)-B_{\text{target}}(z)]e^{i[\theta(z)+qz]}dz,
\end{equation}
where the integral becomes small since $\theta'(z)\approx-q$, and the distance to the target PMF,
\begin{equation}
    |\delta \phi_{\text{L}}(q)| \approx |B(L)-B_{\text{target}}(L)|,
\end{equation}
is minimized, for any $q$, as $|B-B_{\text{target}}|$ is minimized.

\subsection{Refinement: Weighting Function}\label{app:weighting_function}
In practice, the weighting function $\xi(\Delta k)$ in equation~(\ref{eq:objective}) is defined by considering the composition of two windows of $\Delta k$ values: a narrower ``design'' window centered around $\Delta k_0$ with uniform weights $\xi = 1$ (gray area in figure~\ref{fig:refinement}), and a wider window with smaller weights $\xi = 0.1$ used to suppress the background at larger $\left|\Delta k - \Delta k_0 \right|$ values.
The narrower window is set in order to contain a certain fraction of the total target power, $P_{\text{target}} = \int_{\Delta k} \left|\phi_{\text{target}}(q)\right|^2 \, \mathrm{d}q$, for example \SI{99}{\%} of $P_{\text{target}}$, whereas the broader window can be defined in terms of the narrower one, for example, using twice its range.
Within these windows, a grid of $N_q$ phase mismatch values is sampled uniformly, using at least $4$ samples for every $2\pi/L$ interval of $q$.

\subsection{Refinement: step-by-step procedure}\label{app:refinement_procedure}

\begin{enumerate}
    \item Start from the tracked structure and define the narrow and wide phase mismatch windows, as well as the value of $N_q$; examples: narrow window containing \SI{99}{\%} of the target power, wide window twice the narrow window, $N_q=1201$.
    \item Run over the walls and, for each wall, try displacements $\pm k \cdot l_{\text{res}}$, with $k \in \{1, 2, 3, 5, 8, 13, 21\}$; skip displacements that violate the $l_{\text{min}}$ constraint.
    \item Accept the best displacement if it reduces $J$ by more than the threshold.
    \item Evaluate the effect of deleting adjacent wall pairs.
    \item Repeat until a full iteration leaves the structure unchanged.
    \item Check the result on a grid of twice as many frequencies.
\end{enumerate} 

\subsection{Refinement: Simulation results}\label{app:refinement_results}

We examine here the refinement of the two designs using \SI{30}{mm} crystals (OCT default, as discussed in section~\ref{sec:bright_broad_source} and OCT demodulated, as discussed in section~\ref{sec:demod}) and two designs using \SI{5}{mm} crystals (Single-Gaussian PMF and a PMF with a comb of four equidistant Gaussians). Each design is refined against the ideal target PMF (see~\ref{app:target_deviation}).
Besides the objective $J$ of equation~(\ref{eq:objective}), normalized to the weighted target power, $J_{\text{rel}}$, we report the results in terms of the infidelity
\begin{equation}
    1-F = 1-\frac{\left|\sum_{\Delta k}\phi_{\text{target}}^{*}(\Delta k)\,\phi(\Delta k)\right|^2}{\sum_{\Delta k}\left|\phi_{\text{target}}(\Delta k)\right|^2\,\sum_{\Delta k}\left|\phi(\Delta k)\right|^2},
\end{equation}
which is insensitive to the overall scale of $\phi$, and the power ratio
\begin{equation}
    \frac{P}{P_{\text{target}}} = \frac{\sum_{\Delta k}|\phi|^2}{\sum_{\Delta k}|\phi_{\text{target}}|^2},    
\end{equation}
which give the complementary scale value, i.e., the brightness.
Both metrics are evaluated inside the design window, and all the obtained results are summarized in Table~\ref{tab:refinement}.

\begin{table}[htb!]
\caption{\label{tab:refinement}Refinement of the designs:  OCT default (section~\ref{sec:bright_broad_source}) and demodulated (section~\ref{sec:demod}) using ($L=\SI{30}{mm}$, $l_{\text{min}}=\SI{4}{\micro m}$ and $l_{\text{res}}=\SI{125}{nm}$).
All values are evaluated on a validation grid with twice as many sampling points as used by the optimizer.
The single Gaussian and the four-Gaussian PMFs are designed with $L=\SI{5}{mm}$, envelope $\sigma_{\chi}=L/6.04$ (see reference~\cite{faleo2025}), and $l_{\text{c}}=\SI{23}{\micro m}$.}
\footnotesize
\centering
\begin{tabular}{@{}llllllll@{}}
\hline
Design & & $J_{\text{rel}}$ & $1-F$ & $P/P_{\text{target}}$ & Domains & Iterations & Time (s)\\
\hline
OCT, default & tracked & $2.1\times10^{-2}$ & $1.9\times10^{-2}$ & 0.979 & 3481 &  & \\
 & refined & $4.5\times10^{-3}$ & $2.4\times10^{-3}$ & 0.943 & 3469 & 36 & 25\\
OCT, demodulated & tracked & $5.4\times10^{-3}$ & $3.6\times10^{-3}$ & 1.003 & 3619 &  & \\
 & refined & $1.7\times10^{-3}$ & $4.9\times10^{-4}$ & 0.978 & 3619 & 41 & 30\\
Gaussian & tracked & $3.3\times10^{-5}$ & $3.1\times10^{-5}$ & 1.000 & 804 &  & \\
 & refined & $5.2\times10^{-6}$ & $3.5\times10^{-6}$ & 1.000 & 804 & 19 & 1\\
Four Gaussians & tracked & $3.3\times10^{-3}$ & $3.0\times10^{-3}$ & 1.004 & 1017 &  & \\
 & refined & $9.8\times10^{-5}$ & $4.6\times10^{-5}$ & 0.999 & 1017 & 26 & 1\\

\hline
\end{tabular}
\end{table}

\begin{figure*}[htb!]
   \centering
    \includegraphics{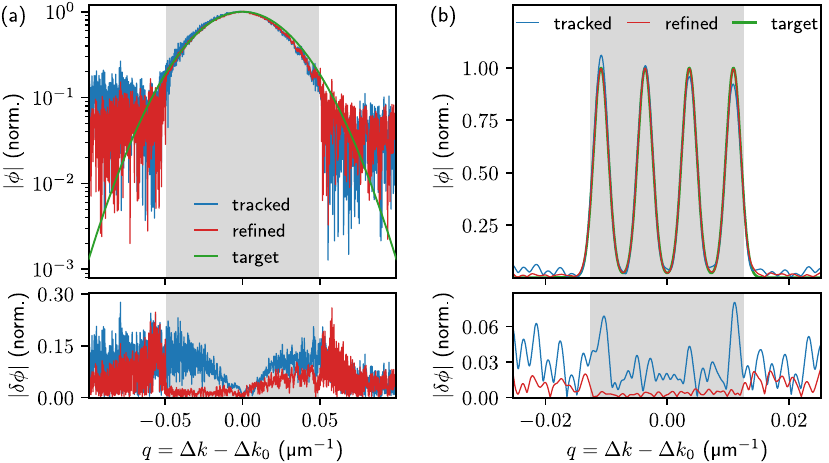}
 \caption{\label{fig:refinement}Refinement of the  designs. The panels compare $\phi(q)$ of the tracked and refined structures against $\phi_{\text{target}}(q)$, as a function of the detuning $q$, for (a) the OCT default design and (b) a design targeting four Gaussians of equal height in a \SI{5}{mm} crystal.
 Upper panels: PMF amplitude.
 Lower panels: complex deviation $|\delta\phi|=|\phi-\phi_{\text{target}}|$.
 All amplitudes are normalized to the maximum of the target.
 The grey shaded area corresponds to uniform weights $\xi=1$, the remaining plotted range to uniform weights $\xi=0.1$ (details in~\ref{app:weighting_function}).}
\end{figure*}

For the OCT design, shown in figure~\ref{fig:refinement}(a), refinement reduces $J_{\text{rel}}$ by a factor of about 5 and the infidelity within the design window by a factor of about 8. The deviation from the target PMF is lower on average over the whole range of detunings $q$ (see lower insets).
The number of domains is almost unchanged.
Starting from the demodulated tracking gives the lowest final error ($J_{\text{rel}}=1.7\times10^{-3}$ against $4.5\times10^{-3}$ for the default tracking).
The lower error comes with a reduction of the power in the design window of about \SI{2.4}{\%} to \SI{3.6}{\%}.
Lower spectral error therefore does not imply unchanged brightness, and the weights $\xi(\Delta k)$ or widths of the detuning windows balance the two, potentially necessitating some tuning according to the application.

A target with several spectral features benefits more from the use of refinement.
We consider here a comb of four Gaussians of equal height, obtained from a Gaussian envelope of width $\sigma_{\chi}=L/6.04$ modulated by $\cos(a\,\Delta z)+\cos(3a\,\Delta z)$, with $\Delta z=z-L/2$ and $a=3/\sigma_{\chi}$ (normalized to $\max|g_{\text{target}}|=2/\pi$).
The PMF consists of four Gaussians centred at $q=\pm a$ and $q=\pm 3a$, as shown in figure~\ref{fig:refinement}(b).

For $L=\SI{5}{mm}$ and $l_{\text{c}}=\SI{23}{\micro m}$, the tracked structure reproduces the peaks with heights decreasing for increasing $q$, with heights of $1.06$, $1.02$, $0.96$ and $0.93$ times the target.

The outer peaks deviate the most, consistently with the expected $q$-weighting of the tracking error in equation~(\ref{eq:error_q}).
Refinement levels all four peaks within \SI{0.5}{\%} from the target, reducing $J_{\text{rel}}$ by a factor 34 and the infidelity by a factor of 65.
The power in the design window remains within \SI{0.2}{\%} of the target power, the number of domains is unchanged, and the refinement converges in 26 iterations, taking less than \SI{2}{\s} using a 13th Gen Intel(R) Core(TM) i5-13500 (2.50 GHz), with (Numba) precompilation excluded.
The large tracking error of the comb therefore stems from its multiple spectral features, and it is this error that the refinement optimization helps removing.

\subsection{Refinement: Efficiency}\label{app:efficiency}
In the refinement procedure, moving a wall $j$ of the $N$ internal walls of a distance $d$, from $z \to z+d$, changes the PMF at every $\Delta k$ of
\begin{equation}
    \delta_j \phi (\Delta k) = 2 s \int_z^{z+d} e^{i \Delta k \bar{z}} \, \mathrm{d}\bar{z},
\end{equation}
with $s=\pm 1$ the sign of the domain on the left of $z$, and where the coefficient 2 comes from the fact that within $d$ the sign of the discrete nonlinearity profile changes from $-s$ to $+s$, i.e., $\Delta g = 2 s$.
By substituting $\bar{z} = z + u$,
\begin{equation}
    \delta_j \phi (\Delta k) = 2 s e^{i \Delta k z} \int_0^{d} e^{i \Delta k u} \, \mathrm{d}u = 2 s e^{i \Delta k z} \cdot I(d,\Delta k),
\end{equation}
the contribution can be separated into two components, as shown in the last equality, which correspond to a phase term depending only on the current wall position $z$ and an integral $I(d,\Delta k)$ depending only on the displacement $d$, for each phase mismatch value.
Given a set of $M$ allowed displacements, for example, $d=\pm k \cdot l_{\text{res}}$, with $k \in \{1, 2, 3, 5, 8, 13, 21\}$, the algorithm computes $I(d,\Delta k)$ only once for all combinations of $d$ values and $N_q$ sampled phase mismatches ($M \times N_q$).
During the evaluation of each wall's displacement, the phase term is computed once per wall per iteration, and used for all candidate displacements.
Considering all $N$ internal walls, the number of operations is proportional to $M \times N_q \times N$ per iteration, with the operations essentially consisting of multiplications and additions.

\end{document}